\documentclass[a4paper,aps,onecolumn]{revtex4}
\RequirePackage[english]{babel}
\RequirePackage[latin1]{inputenc}
\RequirePackage[T1]{fontenc}
\RequirePackage{mathrsfs}
\RequirePackage{amsmath}
\RequirePackage{amssymb}
\RequirePackage{amsbsy}
\RequirePackage{bm}
\def\de#1/de#2{\frac{\partial {#1}}{\partial {#2}}}
\begin{document}
\title{\textbf{Torsion gravity with non--minimally coupled fermionic field: some cosmological models.}}
\author{Stefano Vignolo$^{1}$\footnote{E-mail: vignolo@diptem.unige.it}, Sante Carloni$^{2}$\footnote{E-mail: sante.carloni@tecnico.ulisboa.pt}, Luca Fabbri$^{1}$\footnote{E-mail: fabbri@diptem.unige.it},}
\affiliation{$^{1}$DIME Sez. Metodi e Modelli Matematici, Universit\`{a} di Genova,\\
Piazzale Kennedy, Pad. D, 16129, Genova, ITALY\\
$^{2}$Centro Multidisciplinar de Astrofisica - CENTRA,
Instituto Superior Tecnico - IST,\\
Universidade de Lisboa - UL,
Avenida Rovisco Pais 1, 1049-001, Portugal.}
\date{\today}
\begin{abstract}
We investigate some cosmological models 
arising from a non-minimal coupling of a fermionic field to gravity in the geometrical setting of Einstein-Cartan-Sciama-Kibble gravity. In the presence of torsion, we discuss the role played by  the non-minimal coupling together with fermionic self--interaction potentials in facing problems such as cosmological singularity, inflation and dark energy.
\end{abstract}

\maketitle


\section{Introduction}
In spite of the great successes of General Relativity (GR), cosmological models deriving from Einstein theory of gravity still lack a proper explanation for inflation and dark energy. Inflation and the present cosmic acceleration are indeed two of the main reasons which motivate the study of theories of gravity alternative to General Relativity, at least at large scales.
Among these, scalar--tensor gravitational theories are ones of the most widely investigated already for the beginning 1960s \cite{BD}. Scalar tensor theories also arise in other contexts like the low energy limit of Kaluza-Klein gravity \cite{Overduin:1998pn}, in quantum field theory in curved spacetimes \cite{QFT-CS} and in the tree level action of string theory \cite{Strings}. The basic paradigm of such theories is the non--minimal coupling of gravity to a scalar field, whose so far unknown nature is still the subject of intense scientific research. Among the others, one of the suggested hypotheses is that the scalar field is not fundamental but it is costituted by a fermion condensate. The idea that scalar fields can be composed by other fields (for example Dirac fields) is not very new; for instance, in particle physics it was already proposed  by Weinberg with specific reference to the Higgs field \cite{W} (in this regard see also \cite{Fabbri1,Fabbri2}). 

In cosmology fermion fields have been mostly considered since the 1990s; they have been studied as possible sources of inflation and dark energy, driving the universe into accelerated expansions at both early and late time \cite{Kovalyov,Obukhov,Saha1,Binetruy,Saha2,Picon,Saha3,Ribas1,Saha4,Boehmer1,Boehmer2,Ribas2,Rakhi,Rakhi2,Ribas4,Ribas3,Saha14}.
In most of the papers appeared in the literature fermions are minimally coupled to gravity; only a few and quite recent works instead investigate the effects of fermionic non--minimal couplings \cite{Souza,Devecchi,CVC,GSK}. 

In this paper, we explore some cosmological scenarios when a Dirac field is non--miminally coupled to gravity with torsion. The non--minimal interaction term we take into account is of mass dimension 5 and reduces to the product $\bar\psi\psi\/R$ between the condensate $\bar\psi\psi$ of the Dirac field and the scalar curvature $R$. In a previous paper \cite{FVC}, we have studied the consequences of this non--minimal coupling on the renormalizability of the Dirac equations, showing that in the case torsion is not neglected fermionic non--minimal couplings are renormalizable and possess a well defined behaviour even in the ultraviolet regimes. {  In the present work, we investigate the cosmological counterpart of the theory proposed in \cite{FVC}. Accordingly, here we work within the geometrical setting of Einstein--Cartan--Sciama--Kibble gravity (ECSK), where curvature and torsion couple to energy and spin of the Dirac field respectively. As we shall see, in this metric--affine approach we obtain a 
dynamical equation for the scale volume of the universe which is easier to handle than the analogous one in the purely metric case (compare with \cite{CVC}). As a result, we present a simple analysis of cosmological issues such as cosmological singularity, inflation and dark energy when the above mentioned non--minimal coupling is taken into account. In particular, with the help of some illustrative examples, we discuss the role played by different self--interaction potentials in driving inflation and dark eras in connection with the non--minimal coupling}.
 
The layout of the paper is the following: in section II we briefly outline the theory introduced in \cite{FVC}, recalling its main features; in section III we analyze different cosmological scenarios, first in the presence of a  Dirac field only, then when dust and radiation fluid are present too; finally, we devote section V to the conclusions. Throughout this paper natural units ($\hbar=c=k_{B}=8\pi G=1$) and metric signature $(+,-,-,-)$ are used.


\section{The $(1+\epsilon\bar\psi\psi)R$-theory with torsion}
In this section we briefly review the theory introduced in \cite{FVC}. In the general framework of ECSK gravity, let us consider a Lagrangian density of the form
\begin{equation}\label{3.1}
{\cal L}= (1+\epsilon\bar\psi\psi)\/eR - e\/L_D
\end{equation}
where the Einstein--Hilbert term $R$ is non--minimally coupled to a Dirac Lagrangian of the form
\begin{equation}\label{3.2}
L_D = + \frac{i}{2}\left( \bar{\psi}\Gamma^iD_i\psi - D_i\bar{\psi}\Gamma^i\psi\right) -m\bar{\psi}\psi + V(\bar\psi\psi)
\end{equation}
through the non--minimal coupling term $\epsilon\bar\psi\psi\/R$, $\epsilon$ being a suitable coupling constant. Here, we denote by $\gamma^\mu$ ($\mu=0,1,2,3$) Dirac matrices and we introduce the notation $\Gamma^i = e^i_\mu\gamma^\mu$ where $e^\mu_i$ indicate a tetrad field associated with a metric $g_{ij}=e^\mu_{i}e^\nu_{j}\eta_{\mu\nu}$. In eq. \eqref{3.2}, $D_i$ denote the covariant derivative of the Dirac field $\psi$ defined as 
$D_i\psi = \de\psi/de{x^i} + \omega_i^{\;\;\mu\nu}S_{\mu\nu}\psi\/$ and $D_i\bar\psi = \de{\bar\psi}/de{x^i} - \bar\psi\omega_i^{\;\;\mu\nu}S_{\mu\nu}\/$, where $\omega_i^{\;\;\mu\nu}$ is a spin connection and $S_{\mu\nu}:= \frac{1}{8}[\gamma_\mu,\gamma_\nu]$. Equivalently, we have $D_i\psi = \de\psi/de{x^i} - \Omega_i\psi$ and $D_i\bar\psi = \de{\bar\psi}/de{x^i} + \bar{\psi}\Omega_i$ where
\begin{equation}\label{3.3}
\Omega_i := - \frac{1}{4}g_{jh}\left\{\Gamma_{ik}^{\;\;\;j} - e^j_\mu\partial_i\/e^\mu_k \right\}\Gamma^h\Gamma^k
\end{equation}
and $\Gamma_{ik}^{\;\;\;j}$ are the coefficients of a linear connection $\Gamma$, associated with the spin connection through the usual relation
\begin{equation}\label{3.4}
\Gamma_{ij}^{\;\;\;h} = \omega_{i\;\;\;\nu}^{\;\;\mu}e_\mu^h\/e^\nu_j + e^{h}_{\mu}\partial_{i}e^{\mu}_{j}
\end{equation}
Denoting by $\varphi:=(1+\epsilon\bar\psi\psi)$ and by $V':=\frac{dV}{d(\bar\psi\psi)}$, from \eqref{3.1} we can derive field equations of the form
\begin{subequations}\label{3.9}
\begin{equation}\label{3.9a}
R_{ij} -\frac{1}{2}Rg_{ij}= \frac{1}{\varphi}\Sigma_{ij} 
\end{equation}
\begin{equation}\label{3.9b}
T_{ij}^{\;\;\;h} = - \frac{1}{2\varphi}\de{\varphi}/de{x^p}\left(\delta^p_i\delta^h_j - \delta^p_j\delta^h_i\right) + \frac{1}{\varphi}S^{\;\;\;h}_{ij}
\end{equation}
\begin{equation}\label{3.9c}
i\Gamma^{h}D_{h}\psi + \frac{i}{2}T_h\Gamma^h\psi- m\psi + V'(\bar\psi\psi)\psi - \epsilon\psi\/R=0
\end{equation}
\end{subequations}
where
\begin{equation}\label{3.10}
\begin{split}
\Sigma_{ij} := \frac{i}{4}\/\left( \bar\psi\Gamma_{i}{D}_{j}\psi - {D}_{j}\bar{\psi}\Gamma_{i}\psi \right) -\frac{1}{2}L_D\,g_{ij} =\\
\frac{i}{4}\/\left( \bar\psi\Gamma_{i}{D}_{j}\psi - {D}_{j}\bar{\psi}\Gamma_{i}\psi \right) - \frac{1}{2}\epsilon\bar\psi\psi\/R\,g_{ij} - \frac{1}{2}V(\bar\psi\psi)\,g_{ij} + \frac{1}{2}\bar\psi\psi\/V'(\bar\psi\psi)\,g_{ij}
\end{split}
\end{equation}
and
\begin{equation}\label{3.6}
S_{ij}^{\;\;\;h}=\frac{i}{2}\bar\psi\left\{\Gamma^{h},S_{ij}\right\}\psi
\end{equation}
are respectively the energy--momentum and the spin density tensors. In eqs. \eqref{3.9b} and \eqref{3.9c} $T_{ij}^{\;\;\;h}:=\Gamma_{ij}^{\;\;\;h} - \Gamma_{ji}^{\;\;\;h}$ denotes the torsion tensor and $T_i:=T_{ij}^{\;\;\;j}$ its contraction, while in eq. \eqref{3.6} $S_{ij}:= \frac{1}{8}[\Gamma_i,\Gamma_j]$. The energy--momentum and spin tensors satisfy the conservation laws
\begin{subequations}\label{c.2.6}
\begin{equation}\label{c.2.6a}
\nabla_i\Sigma^{ij} + T_i\Sigma^{ij} - \Sigma_{pq}T^{jpq} -\frac{1}{2}S_{pqr}R^{pqrj} + \frac{1}{2}R\nabla^j\/\varphi= 0
\end{equation}
\begin{equation}\label{c.2.6b}
\nabla_h\/S^{ijh} + T_h\/S^{ijh} + \Sigma^{ij} - \Sigma^{ji} = 0 
\end{equation}
\end{subequations}
automatically ensured by the Dirac equations \eqref{3.9c} \cite{FVC}. It is seen that the antisymmetric part of the Einstein--like equations \eqref{3.9a} amounts to the conservation law for the spin \eqref{c.2.6b}. The significant part of the Einstein--like equations is then the symmetric one which, making use of the trace of \eqref{3.9a} and separating the purely metric terms from the torsional ones through eq. \eqref{3.9b}, can be written in the final form
\begin{equation}\label{3.19}
\begin{split}
\tilde{R}_{ij} -\frac{1}{2}\tilde{R}g_{ij}= \frac{1}{\varphi}\tilde{\Sigma}_{ij}
+ \frac{1}{\varphi^2}\left( - \frac{3}{2}\de\varphi/de{x^i}\de\varphi/de{x^j} + \varphi\tilde{\nabla}_{j}\de\varphi/de{x^i} + 
\frac{3}{4}\de\varphi/de{x^h}\de\varphi/de{x^k}g^{hk}g_{ij} \right. \\
\left. - \varphi\tilde{\nabla}^h\de\varphi/de{x^h}g_{ij}\right) + \frac{3}{64\varphi^2}(\bar{\psi}\gamma_5\gamma^\tau\psi)(\bar{\psi}\gamma_5\gamma_\tau\psi)g_{ij} \\
- \frac{\epsilon(\bar\psi\psi)\left(\frac{m}{2}\bar\psi\psi -2V + \frac{3}{2}\bar\psi\psi\/V'\right)}{2\varphi\left(\frac{1}{2}\varphi - \frac{3}{2}\right)}\,g_{ij} - \frac{1}{2\varphi}V(\bar\psi\psi)\,g_{ij} + \frac{1}{2\varphi}\bar\psi\psi\/V'(\bar\psi\psi)\,g_{ij}
\end{split}
\end{equation}
where $\tilde{R}_{ij}$, $\tilde R$ and $\tilde{\nabla}_i$ are respectively the Ricci tensor, the Ricci scalar curvature and the covariant derivative induced by the Levi--Civita connection and 
\begin{equation}\label{3.15}
\tilde{\Sigma}_{ij} := \frac{i}{4}\/\left[ \bar\psi\Gamma_{(i}\tilde{D}_{j)}\psi - \left(\tilde{D}_{(j}\bar\psi\right)\Gamma_{i)}\psi \right]
\end{equation}
$\tilde{D}_i$ denoting spinor covariant derivative with respect to the Levi--Civita connection. 
The Dirac equations can be handled in a similar way, assuming the expression
\begin{equation}\label{3.20}
i\Gamma^{h}\tilde{D}_{h}\psi
- \frac{1}{\varphi}\frac{3}{16}\left[(\bar{\psi}\psi)
+i(i\bar{\psi}\gamma_5\psi)\gamma_5\right]\psi-m\psi + V'(\bar\psi\psi)\psi - \epsilon\psi\/R=0
\end{equation}
For further details, the reader is referred to \cite{FVC}.

\section{Bianchi--I cosmological models}
\subsection{Coupling to Dirac field only}
{  In order to investigate cosmological scenarios deriving from \eqref{3.19} and \eqref{3.20}}, let us consider a Bianchi type I metric of the form
\begin{equation}\label{4.1}
ds^2 = dt^2 - a^2(t)\,dx^2 - b^2(t)\,dy^2 - c^2(t)\,dz^2
\end{equation}
Denoting by $\tau := abc$ the scale volume, evaluating the linear and spin connection coefficients associated with the metric tensor \eqref{4.1} and inserting the results together with \eqref{4.1} itself in equations \eqref{3.19}, the latter are seen to assume the form
\begin{subequations}\label{4.10}
\begin{equation}\label{4.10a}
\frac{\dot a}{a}\frac{\dot b}{b} + \frac{\dot b}{b}\frac{\dot c}{c} + \frac{\dot a}{a}\frac{\dot c}{c} =
\frac{1}{2\varphi}m\bar\psi\psi - \frac{3}{64\varphi^2}(\bar{\psi}\gamma_5\gamma^\nu\psi)(\bar{\psi}\gamma_5\gamma_\nu\psi) + 
\frac{1}{\varphi^2}\left[- \frac{3}{4}{\dot\varphi}^2 - \varphi\dot\varphi\frac{\dot\tau}{\tau}\right] - \frac{1}{2\varphi}V(\bar\psi\psi) 
\end{equation}
\begin{equation}\label{4.10b}
\begin{split}
\frac{\ddot b}{b} + \frac{\ddot c}{c} + \frac{\dot b}{b}\frac{\dot c}{c} = 
\frac{1}{\varphi^2}\left[\varphi\dot\varphi\frac{\dot a}{a} + \frac{3}{4}{\dot\varphi}^2 -\varphi\left( \ddot\varphi + \frac{\dot\tau}{\tau}\dot\varphi \right)\right] + \frac{3}{64\varphi^2}(\bar{\psi}\gamma_5\gamma^\nu\psi)(\bar{\psi}\gamma_5\gamma_\nu\psi)\\
- \frac{\epsilon(\bar\psi\psi)\left(\frac{m}{2}\bar\psi\psi -2V + \frac{3}{2}\bar\psi\psi\/V'\right)}{2\varphi\left(\frac{1}{2}\varphi - \frac{3}{2}\right)} - \frac{1}{2\varphi}V(\bar\psi\psi) + \frac{1}{2\varphi}(\bar\psi\psi)V'(\bar\psi\psi)
\end{split}
\end{equation}
\begin{equation}\label{4.10c}
\begin{split}
\frac{\ddot a}{a} + \frac{\ddot c}{c} + \frac{\dot a}{a}\frac{\dot c}{c} =  
\frac{1}{\varphi^2}\left[\varphi\dot\varphi\frac{\dot b}{b} + \frac{3}{4}{\dot\varphi}^2 -\varphi\left( \ddot\varphi + \frac{\dot\tau}{\tau}\dot\varphi \right) \right] + \frac{3}{64\varphi^2}(\bar{\psi}\gamma_5\gamma^\nu\psi)(\bar{\psi}\gamma_5\gamma_\nu\psi)\\
- \frac{\epsilon(\bar\psi\psi)\left(\frac{m}{2}\bar\psi\psi -2V + \frac{3}{2}\bar\psi\psi\/V'\right)}{2\varphi\left(\frac{1}{2}\varphi - \frac{3}{2}\right)} - \frac{1}{2\varphi}V(\bar\psi\psi) + \frac{1}{2\varphi}(\bar\psi\psi)V'(\bar\psi\psi)
\end{split}
\end{equation}
\begin{equation}\label{4.10d}
\begin{split}
\frac{\ddot a}{a} + \frac{\ddot b}{b} + \frac{\dot a}{a}\frac{\dot b}{b} =  
\frac{1}{\varphi^2}\left[\varphi\dot\varphi\frac{\dot c}{c} + \frac{3}{4}{\dot\varphi}^2 -\varphi\left( \ddot\varphi + \frac{\dot\tau}{\tau}\dot\varphi \right) \right] + \frac{3}{64\varphi^2}(\bar{\psi}\gamma_5\gamma^\nu\psi)(\bar{\psi}\gamma_5\gamma_\nu\psi)\\
- \frac{\epsilon(\bar\psi\psi)\left(\frac{m}{2}\bar\psi\psi -2V + \frac{3}{2}\bar\psi\psi\/V'\right)}{2\varphi\left(\frac{1}{2}\varphi - \frac{3}{2}\right)} - \frac{1}{2\varphi}V(\bar\psi\psi) + \frac{1}{2\varphi}(\bar\psi\psi)V'(\bar\psi\psi)
\end{split}
\end{equation}
\end{subequations}
together with the conditions
\begin{subequations}\label{4.11}
\begin{equation}\label{4.11a}
\tilde\Sigma_{12}=0\quad \Rightarrow \quad a\/\dot{b} - b\/\dot{a}=0 \quad \cup \quad \bar\psi\gamma^5\gamma^3\psi =0
\end{equation}
\begin{equation}\label{4.11b}
\tilde\Sigma_{23}=0\quad \Rightarrow \quad c\/\dot{b} - b\/\dot{c}=0 \quad \cup \quad \bar\psi\gamma^5\gamma^1\psi =0
\end{equation}
\begin{equation}\label{4.11c}
\tilde\Sigma_{13}=0\quad \Rightarrow \quad a\/\dot{c} - c\/\dot{a}=0 \quad \cup \quad \bar\psi\gamma^5\gamma^2\psi =0
\end{equation}
\end{subequations}
The equations $\tilde\Sigma_{0A}=0$ ($A=1,2,3$) result to be identities. Conditions \eqref{4.11} are constraints imposed on the metric or on the Dirac field. There are three ways to satisfy these conditions: one is to impose constraints of purely geometrical origin by requiring that $a\dot{b}-b\dot{a}=0$, $a\dot{c}-c\dot{a}=0$, $c\dot{b}-b\dot{c}=0$ obtaining an isotropic universe; another is to impose constraints of purely material origin by insisting that $\bar\psi\gamma^5\gamma^1\psi=0$, $\bar\psi\gamma^5\gamma^2\psi=0$, $\bar\psi\gamma^5\gamma^3\psi=0$ giving an anisotropic universe without spin--torsion interactions (in fact in this case necessarily we have that $\bar\psi\gamma^5\gamma^0\psi =0$, otherwise the condition $\bar\psi\gamma^0\psi =0$ must be true, implying that the whole spinor must vanish); the last situation would be of both geometrical and material origin by insisting that for instance $a\dot{b}-b\dot{a}=0$ with $\bar\psi\gamma^5\gamma^1\psi=0$, $\bar\psi\gamma^5\gamma^2\psi=0$ giving a partial isotropy for only two axes with the corresponding two components of the spin vector vanishing. We will be back to this issue in a moment.

Following a useful procedure \cite{Saha1,Saha2,Saha3,VFC,FVC}, we can suitably combine eqs. \eqref{3.19}, obtaining the expressions of the scale factors as functions of the scale volume $\tau$
\begin{subequations}\label{4.13bis}
\begin{equation}\label{4.13bisa}
a= \tau^{\frac{1}{3}}\left(XY\right)^{\frac{1}{3}}e^{\left(\frac{Z+W}{3}\int{\frac{dt}{\varphi\tau}}\right)}
\end{equation}
\begin{equation}\label{4.13bisb}
b=\tau^{\frac{1}{3}}X^{-\frac{2}{3}}Y^{\frac{1}{3}}e^{\left(\frac{-2Z+W}{3}\int{\frac{dt}{\varphi\tau}}\right)}
\end{equation}
\begin{equation}\label{4.13bisc}
c=\tau^{\frac{1}{3}}X^{\frac{1}{3}}Y^{-\frac{2}{3}}e^{\left(\frac{Z-2W}{3}\int{\frac{dt}{\varphi\tau}}\right)}
\end{equation}
\end{subequations}
($X,Y,Z$ and $W$ being integration constants) and the dynamical equation \footnote{ We notice that eq. \eqref{4.14} differs from the analogous equation deduced in \cite{FVC}, due to a trivial calculation error made in \cite{FVC} and responsible for a different expression before the term $V'$; as most of the results obtained in A concern the case $V = 0$, they remain equally exact; the only case discussed in \cite{FVC} where $V\not =0$ remain qualitatively correct.} for $\tau$
\begin{equation}\label{4.14}
2\frac{\ddot\tau}{\tau}  = - 3\frac{\ddot\varphi}{\varphi} - 5\frac{\dot\tau}{\tau}\frac{\dot\varphi}{\varphi} 
- \frac{3m\bar\psi\psi -3V\left(\varphi +1\right) + 3\varphi\bar\psi\psi\/V'}{\varphi\left(\varphi - 3\right)}
\end{equation}
Here, it is noteworthy
that equation \eqref{4.10a} plays the role of a constraint on the initial data and thus on the integration constants. In this regard, in \cite{FVC} it has been actually checked that if the Hamiltonian constraint \eqref{4.10a}
is satisfied initially, then it is preserved in time.
Analogously, in the metric \eqref{4.1} the Dirac equations \eqref{3.20} become 
\begin{subequations}\label{4.9}
\begin{equation}\label{4.9a}
\dot\psi + \frac{\dot\tau}{2\tau}\psi + im\gamma^0\psi +
\frac{3i}{16\varphi}\/\left[ (\bar\psi\psi)\gamma^0 +i\/(i\bar\psi\gamma^5\psi)\gamma^0\gamma^5 \right]\psi + i\epsilon\/R\gamma^0\psi -iV'\gamma^0\psi=0
\end{equation}
\begin{equation}\label{4.9b}
\dot{\bar\psi} + \frac{\dot\tau}{2\tau}\bar\psi - im\bar{\psi}\gamma^0 - \frac{3i}{16\varphi}\bar\psi\/\left[ (\bar\psi\psi)\gamma^0 +i\/(i\bar\psi\gamma^5\psi)\gamma^5\gamma^0 \right] - i\epsilon\/R\bar\psi\gamma^0 + iV'\bar\psi\gamma^0=0
\end{equation}
\end{subequations}
{  In order to solve eqs. \eqref{4.9}, we can adapt to the present context the arguments developed in \cite{Saha14}. First of all we combine eqs. \eqref{4.9} obtaining the differential equations
\begin{equation}\label{4.17a}
\frac{d}{dt}\left(\tau\bar\psi\psi\right) + \frac{3\tau}{8\varphi}\left(i\bar\psi\gamma^5\psi\right)\left(\bar\psi\gamma^5\gamma^0\psi\right)=0
\end{equation}
We search for solutions of \eqref{4.9} satisfying the condition
\begin{equation}\label{4.17a.1}
\bar\psi\gamma^5\psi =0
\end{equation} 
Under such a hypothesis, eq. \eqref{4.17a} implies
\begin{equation}\label{4.18}
\bar\psi\psi = \frac{K}{\tau}
\end{equation}
where $K$ is an integration constant. At the same time, from eqs. \eqref{3.9a} and \eqref{3.9c} we can derive the expression of the scalar curvature $R$ as function of the bilinear spinor $\bar\psi\psi$
\begin{equation}\label{4.17a.2}
R(\bar\psi\psi)=\frac{m\bar\psi\psi -4V + 3\bar\psi\psi\/V'}{\varphi -3}
\end{equation}
In view of eqs. \eqref{4.17a.1}, \eqref{4.18} and \eqref{4.17a.2}, the Dirac equation \eqref{4.9a} can be rewritten as
\begin{equation}\label{4.17a.3}
\dot\psi + \frac{\dot\tau}{2\tau}\psi + iG(\tau)\gamma^0\psi =0
\end{equation} 
where we have set
\begin{equation}\label{4.17a.4}
G(\tau) := \left(m + \frac{3}{16\varphi}\bar\psi\psi + \epsilon\/R(\bar\psi\psi) - V'(\bar\psi\psi)\right)_{|\bar\psi\psi = \frac{K}{\tau}}
\end{equation}
Considering the 4-component spinor field 
\begin{eqnarray}\label{4.17a.5}
\psi =\left(\begin{tabular}{c}
$\psi_1$\\
$\psi_2$\\
$\psi_3$\\
$\psi_4$
\end{tabular}\right)
\end{eqnarray}
eqs. \eqref{4.17a.3} assume the explicit form
\begin{subequations}\label{4.17a.6}
\begin{equation}
\dot{\psi}_1 + \frac{\dot\tau}{2\tau}\psi_1 + iG(\tau)\psi_1 =0
\end{equation}
\begin{equation}
\dot{\psi}_2 + \frac{\dot\tau}{2\tau}\psi_2 + iG(\tau)\psi_3 =0
\end{equation}
\begin{equation}
\dot{\psi}_3 + \frac{\dot\tau}{2\tau}\psi_3 - iG(\tau)\psi_3 =0
\end{equation}
\begin{equation}
\dot{\psi}_4 + \frac{\dot\tau}{2\tau}\psi_4 - iG(\tau)\psi_4 =0
\end{equation}
\end{subequations}
A solution of eqs. \eqref{4.17a.6} is then given by 
\begin{eqnarray}\label{4.17a.7}
&\psi=\frac{1}{\sqrt{\tau}}\left(\begin{tabular}{c}
$C_1\exp\left(-i\int{Gdt}\right)$\\
$C_2\exp\left(-i\int{Gdt}\right)$\\
$C_3\exp\left(+i\int{Gdt}\right)$\\
$C_4\exp\left(+i\int{Gdt}\right)$
\end{tabular}\right)
\end{eqnarray}
where $C_i$ are integration constants which, because of constraints \eqref{4.17a.1} and \eqref{4.18}, have to satisfy the equations
\begin{subequations}\label{4.17a.8}
\begin{equation}
C_1^*C_1 + C_2^*C_2 - C_3^*C_3 - C_4^*C_4 = K\label{a}
\end{equation}
\begin{equation}
C_1^*C_3 + C_2^*C_4 =0\label{b}
\end{equation}
\end{subequations}
Moreover, the constants $C_i$ have to satisfy further constraints deriving from the non diagonal part of the Einstein--like equations \eqref{3.19}. As we have discussed above, these additional conditions result in a maximum of three real equations given by
\begin{subequations}\label{4.17a.9}
\begin{equation}
C_1^*C_2 + C_2^*C_1 + C_3^*C_4 + C_4^*C_3 =0
\end{equation} 
\begin{equation}
C_1^*C_2 - C_2^*C_1 + C_3^*C_4 - C_4^*C_3 =0
\end{equation} 
\begin{equation}
- C_1^*C_1 + C_2^*C_2 - C_3^*C_3 + C_4^*C_4 =0
\end{equation} 
\end{subequations}
Equations \eqref{4.17a.8} and \eqref{4.17a.9} form a system of at most six real equations for eight real unknowns, thus in general one should expect that solutions exist. 

However, if all of the last constraints were considered then we can draw some additional conclusions. By combining the first two of \eqref{4.17a.9} we see that 
\begin{equation}
C_1^*C_2+C_3^*C_4=0\label{aux}
\end{equation}
which can be combined together with the second of \eqref{4.17a.8} to show that 
\begin{equation}
C_2(|C_1|^2-|C_4|^2)= C_2C_1^*C_1-C_2C_4^*C_4=-C_4C_3^*C_1+C_3^*C_1C_4=0
\end{equation} 
and so either $C_2=0$ or $|C_1|^2=|C_4|^2$ in general. If $|C_1|^2=|C_4|^2$  we would have that the third of \eqref{4.17a.9} remains $|C_2|^2=|C_3|^2$ and so $K=0$. If $C_2=0$ we have that \eqref{aux} and the second of \eqref{4.17a.8}  imply that either $C_3=0$ or $C_1=C_4=0$, which by the \eqref{4.17a.9} yields again $C_3=0$. In both cases this returns again $K=0$. Therefore, if all \eqref{4.17a.9} are accounted for, then $K$ is necessarily zero, and there would be no condensate: this is to be expected, because these three conditions are equivalent to the requirement of total isotropy of the spinor field. Indeed, if this were to be the case, then  all of the spatial components of the spin vector would have to vanish, and because the algebraic identity $\bar\psi\gamma^5\gamma^\mu\psi\bar\psi\gamma_\mu\psi=0$ is always true, then  $\bar\psi\gamma^5\gamma^0\psi\bar\psi\gamma_0\psi=0$. Now, if  $\psi\bar\psi\gamma_0\psi=\psi^{\dagger}\psi=0$, the spinor itself would be zero.  Therefore we have to select the condition $\bar\psi\gamma^5\gamma^0\psi =0$, which means that also the temporal component of the spin vector is zero, and therefore the entire spin vector is zero. Furthermore, since  
\begin{equation}
|i\bar\psi\gamma^5\psi|^2+|\bar\psi\psi|^2=-\bar\psi\gamma^5\gamma^\mu\psi\bar\psi\gamma^5\gamma_\mu\psi
\end{equation}
the reasoning above leads the conditions $i\bar\psi\gamma^5 \psi=\bar\psi\psi=0$, which imply that there is no condensate, and thus no non-minimal coupling. In conclusion, we have to dismiss the case  $\bar\psi\gamma^5\gamma^1\psi=0$, $\bar\psi\gamma^5\gamma^2\psi=0$, $\bar\psi\gamma^5\gamma^3\psi=0$ because it is not possible to have a geometrically anisotropic universe filled with isotropic matter. 

A second scenario is to have partial isotropy in both geometry and matter as for instance in the case $a\dot{b}-b\dot{a}=0$ with $\bar\psi\gamma^5\gamma^1\psi=0$, $\bar\psi\gamma^5\gamma^2\psi=0$. In this circumstance the last of \eqref{4.17a.9} is lost and thus a solution can be found. For example, a solution is given by $|C_1|^2=K$ and all other constants equal to zero or by $|C_2|^2=K$ and all other constants equal to zero: they give
\begin{eqnarray}\label{soluzioniesatte}
&\psi=\exp\left(-i\int{Gdt}\right)\sqrt{\frac{K}{\tau}}\left(\begin{tabular}{c}
$1$\\
$0$\\
$0$\\
$0$
\end{tabular}\right)\ \ \ \ \ \ \ \mathrm{or} \ \ \ \ \ \ \ \ \ 
\psi=\exp\left(-i\int{Gdt}\right)\sqrt{\frac{K}{\tau}}\left(\begin{tabular}{c}
$0$\\
$1$\\
$0$\\
$0$
\end{tabular}\right)
\end{eqnarray}
which can respectively be interpreted as a spinor in the spin $\frac{1}{2}$ or the spin $-\frac{1}{2}$ eigenstate in non-relativistic case (that is, with the two lower components vanishing in standard representation).

The third and last case is given by a totally isotropic universe $a\dot{b}-b\dot{a}=0$, $a\dot{c}-c\dot{a}=0$, $c\dot{b}-b\dot{c}=0$ filled with anisotropic matter \footnote{This might seem a contradiction, but the fact hat an anisotropic spinor could be compatible with an isotropic metric is due to the way in which the spinor gravitates. Looking at equations \eqref{4.10} it clear that the spin vector, which would be responsible of the breaking of isotropy enters in the field equations only as $(\bar{\psi}\gamma_5\gamma^\tau\psi)(\bar{\psi}\gamma_5\gamma_\tau\psi)$. Therefore spacetime does not ``respond'' to the spin vector and can be isotropic.}.  In such a circumstance eqs. \eqref{4.17a.9} do not apply and eqs. \eqref{4.17a.8} certainly admit solutions, for instance still of the form \eqref{soluzioniesatte}.

We remark that in any case, the condensate evolves as $\bar\psi\psi=\frac{K}{\tau}$ and that is all we need to perform the analysis of the cosmological model. The fact that the entire information about the spinor is not necessary and that only the condensate is important may sound strange but it is exactly what we would expect to have in macroscopic systems (after all, also in the physics of condensates one does not need the complete dynamical behaviour of each single electron or Cooper couple to know the evolution of the condensate itself --- similar arguments can be used to justify why one does not need the motion of each single atom or molecule to know the evolution of a gas).} 

So, resuming the problem of finding the dynamical equation for the scale volume $\tau$, we may insert the relation $\bar\psi\psi = \frac{K}{\tau}$ into \eqref{4.14} getting the final equation
\begin{equation}\label{1}
2\frac{\ddot\tau}{\tau}\varphi+3\ddot\varphi+5\frac{\dot\tau}{\tau}\dot\varphi
=\frac{3mK}{\tau\left(2-\frac{\epsilon K}{\tau}\right)} - \frac{3\left(\epsilon K + 2\tau\right)\/V}{\tau\left(2-\frac{\epsilon K}{\tau}\right)} + \frac{3\left(\epsilon\/K+\tau\right)KV'}{\tau^2\left(2-\frac{\epsilon\/K}{\tau}\right)}
\end{equation}
The fact that one can combine the equations in this way should not be surprising. In fact, given a time-like normalized vector field $X_a$ and the projection tensor $h_{ab}=g_{ab}-X_aX_b$, the physical properties of any anisotropic cosmology can be characterized by the expansion scalar and and the shear scalar:  
\begin{equation}
\theta=h^{ab}\tilde{\nabla}_{a}X_b \qquad \sigma=\frac{1}{2}\sqrt{\sigma_{ab}\sigma^{ab}} \qquad \sigma_{ab}=\frac{1}{2}h^{c}_{a}h^{d}_{b}\left(\tilde{\nabla}_{c}X_{d}+\tilde{\nabla}_{d}X_{c}\right) -\frac{1}{3}h_{ab}\theta
\end{equation}
In the particular case of the metric \eqref{4.1} the expansion is 
\begin{equation}
\theta=\frac{\dot{a}}{a}+\frac{\dot{b}}{b}+\frac{\dot{c}}{c}=\frac{\dot{\tau}}{\tau}
\end{equation}
so that the \eqref{1} is an analogous of the Raychaudhuri equation. It is also useful to write the shear scalar in terms of the metric \eqref{4.1} ad the $\tau$
\begin{equation}\label{1.sigma}
\sigma=\frac{1}{2}\left[\left(\frac{\dot{a}}{a}\right)^2+\left(\frac{\dot{b}}{b}\right)^2+\left(\frac{\dot{c}}{c}\right)^2- \frac{1}{3}\left(\frac{\dot{a}}{a}+\frac{\dot{b}}{b}+\frac{\dot{c}}{c}\right)^2\right]^{1/2}=\frac{1}{2}\left[\frac{2(Z^2+W^2-ZW)}{3(\tau+\epsilon K)^2}\right]^{1/2}
\end{equation}
where in the last expression we have used the \eqref{4.13bis}. It is immediately clear that the only way to increase the anisotropy of the system is to have a contraction, so these models, if expanding, tend to isotropize. In addition, and differently from GR, for $\tau \rightarrow 0$ (and for $\epsilon >0$) the shear tends to a finite value depending on $K$ and other constants of integration; on the contrary, if $\epsilon < 0$, the shear scalar can blow up before that $\tau =0$.

Using the identity $2\ddot\tau\varphi+3\tau\ddot\varphi+5\dot\tau\dot\varphi = \frac{d^2}{dt^2}\left(2\tau-\epsilon K\ln\tau\right)$, \eqref{1}  yields
\begin{equation}\label{2}
\frac{d}{dt}\left[\frac{d}{dt}\left(2\tau-\epsilon K\ln\tau\right)\right]^{2}
=6\left[mK - \left(\epsilon\/K + 2\tau\right)V + \frac{\left(\epsilon\/K+\tau\right)K}{\tau}V'\right]\dot{\tau}
\end{equation}
In the following, by exploiting the linear dependence on the potential and its derivative in eq. \eqref{1} (or \eqref{2}), we analyze different scenarios associated to various choices of the potential $V$. To do that we follow two different approaches: the first one is a reconstruction technique, where a given time evolution for the scale volume is assumed and then eq. \eqref{1} is solved for $V$, making systematically use of the relation $\bar\psi\psi = \frac{K}{\tau}$; the second one consists in choosing $V$ in such a way that the right--hand side of eq. \eqref{2} becomes easily solvable (exactly or at least for some approximations) and the corresponding solutions represent interesting cosmological evolutions. The properties of these scenarios will be characterised in terms of the behavior of $\theta$ and $\sigma$.


\subsubsection{The case $V=0$.} 
To start with, we discuss the simplest case $V=0$. In this circumstance eq. \eqref{2} assumes the form
\begin{equation}\label{3}
\frac{d}{dt}\left[\frac{d}{dt}\left(2\tau-\epsilon K\ln\tau\right)\right]^{2}=6mK\dot{\tau}
\end{equation}
which can be integrated as
\begin{equation}\label{4}
\frac{d}{dt}(2\tau-\epsilon K\ln{\tau})\!=\!\pm\sqrt{6mK\tau\!-\!A}
\end{equation}
yielding a first--order differential equation for $\tau$ with integration constant $A$. Assuming $A$ be negative, equation \eqref{4} can be integrated as
\begin{equation}\label{5}
t\!+\!B\!=\!\pm\frac{2\sqrt{|A|}}{3mK}\left(\sqrt{\frac{6mK}{|A|}\tau\!+\!1}\right)\!\pm\!
\frac{2\epsilon K}{\sqrt{|A|}}\mathrm{arctanh}\left(\sqrt{\frac{6mK}{|A|}\tau\!+\!1}\right)
\end{equation}
but as it is also clear, $A$ negative (with of course $\tau$ positive) means that the argument of the $\mathrm{arctanh}$ is larger than one and thus such function is ill-defined. Therefore we are forced to assume $A\geq 0$: in the case $A >0$ the differential equation is integrated as
\begin{equation}\label{6}
t\!+\!B\!=\!\pm\frac{2\sqrt{A}}{3mK}\left(\sqrt{\frac{6mK}{A}\tau\!-\!1}\right)\!\mp\!
\frac{2\epsilon K}{\sqrt{A}}\arctan{\left(\sqrt{\frac{6mK}{A}\tau\!-\!1}\right)}
\end{equation}
which is well-defined whenever the volume is larger than a given lower-bound $\tau_{0}\!\geqslant\!\frac{A}{6mK}$ and thus showing that, regardless the value of $B$, there is no way in which the minimal volume $\tau_{0}$ can be zero; if $A=0$, we get the solution
\begin{equation}
t\!+\!B\!=\!\pm\frac{\sqrt{2}\left(\epsilon\/K+2\tau\right)}{\sqrt{3mK\tau}}
\label{zero}
\end{equation}
from which again we cannot have zero scale volume at a finite time. In all these cases then, singularities are avoided due to the presence of the non-minimal coupling term we have here: in fact, if $\epsilon\!=\!0$ then there will be nothing preventing us to have a negative $A$, so that it would be possible to have the solution \eqref{5} which in this case would reduce to
\begin{equation}
t\!+\!B\!=\!\pm\frac{2\sqrt{|A|}}{3mK}\left(\sqrt{\frac{6mK}{|A|}\tau\!+\!1}\right)
\end{equation}
allowing zero scale volume $\tau=0$ at the finite time $t\!=\!-B\pm\frac{2\sqrt{|A|}}{3mK}$. These phenomena are not new in the context of ECSK theories (with minimal \cite{Poplawski:2009su,Poplawski:2011jz,Poplawski:2011wj,Poplawski:2010jv} and non minimal couplings \cite{Magueijo:2012ug}). However, differently form these studies, our analysis relies exclusively on the exact field equations and therefore it is of purely mathematical nature. 

Also, it must be pointed out that the analysis of the Hamiltonian constraint \eqref{4.10a} provides interesting constraints on the constants of this model. To see this point, using  eqs. \eqref{4.13bis} and \eqref{4}, we easily get the identities    
\begin{subequations}\label{0.0.1}
\begin{equation}
\frac{\dot a}{a} = \frac{1}{3}\frac{\dot \tau}{\tau} + \frac{(Z+W)}{3} \frac{1}{\varphi\tau}
\end{equation}
\begin{equation}
\frac{\dot b}{b} = \frac{1}{3}\frac{\dot \tau}{\tau} + \frac{(-2Z+W)}{3} \frac{1}{\varphi\tau}
\end{equation}
\begin{equation}
\frac{\dot c}{c} = \frac{1}{3}\frac{\dot \tau}{\tau} + \frac{(Z-2W)}{3} \frac{1}{\varphi\tau}
\end{equation}
\begin{equation}
{\dot \tau}^2 =\frac{\tau^2}{(2\tau - \epsilon\/K)^2}(6mK\tau -A)
\end{equation}
\end{subequations}
Inserting the content of \eqref{0.0.1} into \eqref{4.10a}, we obtain the relation
\begin{equation}\label{0.0.8}
-\frac{A}{12} + \frac{1}{9}\left[-3(Z+W)^2 + 9ZW\right] = K^2\left(\frac{m}{2}\epsilon + \frac{3}{64}\right)
\end{equation}
Because of the restriction imposed on A ($A\geq 0$) found above, the left hand side of \eqref{0.0.8} is always non--positive and so must be the right hand side: this necessarily requires 
\begin{equation}\label{0.0.9}
\epsilon \leq -\frac{3}{32m}
\end{equation}
which represents an upper bound for the coupling constant $\epsilon$ in the case the self--interaction potential $V$, or also other kinds of matter different from the only fermionic field, are absent. This fact, together with eq. \eqref{1.sigma}, implies that the singularity on the scale factors can be replaced by a singularity in the shear that happens at finite time (if $\frac{A}{6mK}\leq |\epsilon|K$). In this respect, therefore, the claim that these models are singularity free is an incomplete statement, as the model could retain a singularity (albeit of a different type) at some point in its history.

Another interesting aspect associated with the non--minimal coupling we are studying is that if there were a (cosmological) time interval in which the first term on the right hand side of equation \eqref{6} were negligible with respect to the second one, then in such a time interval we would have an expansion of the universe according to $\tau\!\sim\!\left(\tan{t}\right)^{2}$, which could account for an accelerated behaviour possibly fitting inflationary scenarios (at least for isotropic models). The above mentioned circumstance could be achieved for example by assigning initial data and then integration constants such that $\sqrt{A}/K$ is very small.

The model outlined above is therefore rather intriguing, because it can solve the problem of the cosmological singularity in quite elegant a way and simultaneously, by a careful fine tuning, it can address the issue of inflationary scenarios. Unfortunately, the model with $V=0$ is unable to account for cosmic acceleration at late time. This is easily seen still considering equation \eqref{6}, this time evaluated for large values of $\tau$ (with respect to a given reference volume of the universe), obtaining a behaviour of the scale volume as $\tau\!\sim\!t^{2}$ i.e. $\theta=2/t$, which at late time ensures isotropization (see eqs. \eqref{4.13bis}) but under a decelerated expansion of the scale factors. 


\subsubsection{The potential for a decelerated power law expansion.}
As a first example in which a potential is present, following a reconstruction approach we look for a potential $V$ which gives rise to an expansion law of the form $\tau=\tau_0\/t^2$ already treated in the previous section. This behaviour of $\tau$ implies that the scale factors $a, b ,c$ have a decelerated expansion law, at least at late time.

Inserting $\tau=\tau_0\/t^2$ into \eqref{1}, multiplying by $\tau$ and expressing all in terms of $\bar\psi\psi$, we get the differential equation for the unknown $V$ 
\begin{equation}\label{F1}
2\tau_0 + 2\epsilon\tau_0\bar\psi\psi = \frac{3mK}{\left(2-\epsilon\bar\psi\psi\right)} - \frac{3K\left(\epsilon\bar\psi\psi + 2\right)}{\bar\psi\psi\left(2-\epsilon\bar\psi\psi\right)}V + \frac{3K\left(\epsilon\bar\psi\psi+1\right)}{\left(2-\epsilon\bar\psi\psi\right)}V'
\end{equation}
The solution of \eqref{F1} is
\begin{equation}\label{F2}
V(\bar\psi\psi)=\frac{1}{\left(\epsilon\bar\psi\psi+1\right)}\left[-\frac{2\epsilon\tau_0}{3K}\left(\bar\psi\psi\right)^3 - \frac{4\tau_0}{3K}\left(\bar\psi\psi\right) + \frac{2\epsilon\tau_0}{3K}\left(\bar\psi\psi\right)^2\ln\left(\bar\psi\psi\right) + m\left(\bar\psi\psi\right)\right]
\end{equation}
In addition, the cosmology isotropizes ( $\sigma\rightarrow 0$) in the future, since  
\begin{equation}\label{s-V=0}
\sigma=\frac{1}{2}\left[\frac{2(Z^2+W^2-ZW)}{3\left(\tau_0 t^2+\epsilon K\right)^2}\right]^{1/2}
\end{equation} 
As above, this results could be deduced also from the \eqref{4.13bis}, which converge to $a\propto b\propto c$ for this behaviour of $\tau$.


\subsubsection{Potentials for exponential expansion.}
As a second example, we search for potentials inducing exponential expansion of the scale volume. We begin by a reconstruction technique considering a scale volume of the form $\tau=\tau_0\exp(t)$. In this case $\theta=1$, and
\begin{equation}
\sigma=\frac{1}{2}\left\{\frac{2(Z^2+W^2-ZW)}{3\left[\tau_0\exp(t)+\epsilon K\right]^2}\right\}^{1/2}
\end{equation}
so that the anisotropy becomes quickly zero. Inserting $\tau=\tau_0\exp(t)$ into \eqref{1}, multiplying by $\tau$ and using $\bar\psi\psi$ as independent variable, we get the final equation
\begin{equation}\label{E1}
\frac{2K}{\bar\psi\psi} = \frac{3mK}{\left(2-\epsilon\bar\psi\psi\right)} - \frac{3K\left(\epsilon\bar\psi\psi + 2\right)}{\bar\psi\psi\left(2-\epsilon\bar\psi\psi\right)}V + \frac{3K\left(\epsilon\bar\psi\psi+1\right)}{\left(2-\epsilon\bar\psi\psi\right)}V'
\end{equation}
The latter admits the solution
\begin{equation}\label{E2}
V(\bar\psi\psi)=\frac{\bar\psi\psi\left(2\epsilon+3m\right)-2}{3\left(\epsilon\bar\psi\psi+1\right)}
\end{equation}
Another potential which yields exponential expansion at least at late time is given by
\begin{equation}\label{E3}
V(\bar\psi\psi) = - \frac{1}{6\left(\epsilon\bar\psi\psi+1\right)}
\end{equation}
Indeed, with the choice \eqref{E3}, equation \eqref{2} can be integrated as
\begin{equation}\label{ded11}
\frac{\left(2\tau-\epsilon K\right)\dot\tau}{\tau} = \sqrt{6mK\tau + \tau^2 + A}
\end{equation}
with $A$ denoting an integration constant. It is evident that if  $A$ is negative there exists automatically a strictly positive minimum value of the scale volume, then the singularity in the scale volume is avoided. For instance, setting $A=-1$ for simplicity, eq. \eqref{ded11} can be integrated as
\begin{equation}\label{ded11bis}
t+C= 2\ln\left(\sqrt{6mK\tau+\tau^2-1}+ 3mK + \tau\right) - \epsilon\/K\arctan{\left(\frac{3mK\tau - 1}{\sqrt{6mk\tau + \tau^2 -1}}\right)}
\end{equation}
which for large values of $\tau$ yields exponential expansion. We discuss more in detail the case $A>0$; in such a circumstance by integrating eq. \eqref{E3} we get
\begin{equation}\label{ded13}
t+C=2\ln{\left(\sqrt{6mK\tau + \tau^2 + A}+3mK+\tau\right)}\!+\!
\frac{\epsilon K}{\sqrt{A}}\ln\left(\frac{A+3mK\tau+\sqrt{A}\sqrt{6mK\tau + \tau^2 + A}}{\tau}\right) 
\end{equation}
For large values of $\tau$ we have as above exponential expansion of the scale volume; moreover, setting $\epsilon<0$, for very small values of $\tau$ (with respect to a given reference volume of the universe) we can approximate the solution \eqref{ded13} to 
\begin{equation}\label{ded13bis}
t+D = \frac{\epsilon\/K}{\sqrt{A}}\ln{\left(\frac{2A}{\tau}\right)}
\end{equation}
yielding again exponential expansion. We notice that both the potentials \eqref{E2} and \eqref{E3} are not trivial in view of the non--minimal coupling. Indeed, if $\epsilon =0$ \eqref{E3} reduces to a cosmological constant while \eqref{E2} makes the Lagrangian \eqref{3.2} identical to that of a massless Dirac spinor with cosmological constant. 


\subsubsection{Potentials for transition from an early power law inflation to a decelerated power law expansion era.}
Let us consider the potential
\begin{equation}\label{IF1}
V(\bar\psi\psi)= \frac{\gamma\left(\bar\psi\psi\right)^{p+1}}{6K^{p+1}\left(p-1\right)\left(\epsilon\bar\psi\psi+1\right)}
\end{equation}
where $\gamma$ is a suitable constant. It is easily seen that for such choice of potential eq. \eqref{2} becomes
\begin{equation}\label{IF2}
\frac{d}{dt}\left[\frac{d}{dt}\left(2\tau-\epsilon K\ln\tau\right)\right]^{2} = \left(6mK + \gamma\tau^{-p}\right)\dot\tau
\end{equation}
From \eqref{IF2}, by integrating we get
\begin{equation}\label{IF3}
\left(2-\frac{\epsilon\/K}{\tau}\right)^2{\dot\tau}^2 = 6mK\tau + \frac{\gamma}{-p+1}\tau^{-p+1} + A
\end{equation}
$A$ being an integration constant. Now, for large values of $\tau$ eq. \eqref{IF3} approximates the equation
\begin{equation}\label{IF5}
2|\dot\tau| = \sqrt{6mK\tau}
\end{equation}
giving rise to $\tau \approx t^2$ and then to isotropization. On the contrary, for very small values of $\tau$ eq. \eqref{IF3} can be approximated by
\begin{equation}\label{IF6}
|\epsilon|\/K\frac{|\dot\tau|}{\tau}= \sqrt{\frac{\gamma}{1-p}}\tau^{\frac{1-p}{2}}.
\end{equation}
By choosing $p$ such that $\frac{1-p}{2} = -\frac{1}{3q}$, $q\geq 2$ being an even number, we have $\tau \approx t^{3q}$ which represents power law inflation  at least for isotropic models. In the case of initial anisotropy, the shear scalar 
\begin{equation}
\sigma=\frac{1}{2}\left[\frac{2(Z^2+W^2-ZW)}{3\left(\tau_0 t^{3q}+\epsilon K\right)^2}\right]^{1/2}
\end{equation} 
ensures a quick isotropization, depending on $q$. {  It should be noted that potentials of the form \eqref{IF1} would work equally well also in the case of minimal coupling ($\epsilon =0$)}.


\subsubsection{Potentials for transition from a decelerated expansion era to dark era.}
Let us consider the scale volume function of the form 
\begin{equation}\label{FD1}
\tau = \tau_0\left(\sinh\left(\lambda\/t\right)\right)^2
\end{equation}
for which the expansion is $\theta=2 \lambda  \coth (\lambda  t)$ and the shear scalar is
\begin{equation}
\sigma=\frac{1}{2}\left\{\frac{2(Z^2+W^2-ZW)}{ 3\left[\tau_0\sinh^2\left(\lambda\/t\right)+\epsilon K\right]^2}\right\}^{1/2}
\end{equation}
i.e. a cosmology for which there is a transition between a power law and a de Sitter expansion and the anisotropy decreases converging eventually to zero.
In this case, we have the identity
\begin{equation}\label{FD2}
\frac{d^2}{dt^2}\left(2\tau-\epsilon K\ln\tau\right) = 8\lambda^2\tau + 4\tau_0\lambda^2 + \frac{2\epsilon\/K\lambda^2\tau_0}{\tau} = \frac{8\lambda^2\/K}{\bar\psi\psi} + 4\tau_0\lambda^2 + 2\epsilon\lambda^2\/\tau_0\bar\psi\psi
\end{equation}
In view of \eqref{FD2}, eq. \eqref{1} assumes the form
\begin{equation}\label{FD3}
\frac{8\lambda^2\/K}{\bar\psi\psi} + 4\tau_0\lambda^2 + 2\epsilon\lambda^2\/\tau_0\bar\psi\psi = \frac{3mK}{\left(2-\epsilon\bar\psi\psi\right)} - \frac{3K\left(\epsilon\bar\psi\psi + 2\right)}{\bar\psi\psi\left(2-\epsilon\bar\psi\psi\right)}V + \frac{3K\left(\epsilon\bar\psi\psi+1\right)}{\left(2-\epsilon\bar\psi\psi\right)}V'
\end{equation}
A solution of \eqref{FD3} is given by
\begin{equation}\label{FD4}
V(\bar\psi\psi)= - \frac{2\lambda^2\tau_0\bar\psi\psi}{3K\left(\epsilon\bar\psi\psi+1\right)}\left[\epsilon^2\left(\bar\psi\psi\right)^2 + 4\right] + \frac{m\bar\psi\psi}{\epsilon\bar\psi\psi+1} - \frac{8\lambda^2}{3}
\end{equation}
\subsubsection{Potentials for transition power law inflation -- decelerated power law expansion -- dark era}
Now, let us consider a suitable combination of the potentials introduced above as  
\begin{equation}\label{IFD1}
V(\bar\psi\psi) = - \frac{\alpha}{6\left(\epsilon\bar\psi\psi+1\right)} - \frac{\beta\bar\psi\psi}{6K\left(\epsilon\bar\psi\psi+1\right)} + \frac{\gamma\left(\bar\psi\psi\right)^{p+1}}{6K^{p+1}\left(p-1\right)\left(\epsilon\bar\psi\psi+1\right)}
\end{equation}
$\alpha,\beta$ and $\gamma$ being constants. This particular choice of potential gives rise to a dynamical equation for the scale volume of the form
\begin{equation}\label{IFD2}
\frac{d}{dt}\left[\frac{d}{dt}\left(2\tau-\epsilon K\ln\tau\right)\right]^{2} = \left(6mK + 2\alpha\tau + \beta + \gamma\tau^{-p}\right)\dot\tau
\end{equation}
which, integrated a first time, yields
\begin{equation}\label{IFD3}
\left(2-\frac{\epsilon\/K}{\tau}\right)^2{\dot\tau}^2 = \left(6mK+\beta\right)\tau +\alpha\tau^2 + \frac{\gamma}{-p+1}\tau^{-p+1} - A
\end{equation}
$A$ being an integration constant. Choosing $p$ as above, for very small values of $\tau$ we recover a power law inflation phase; for large values of $\tau$ we recover exponential expansion but, by carefully choosing the values of the parameter $\alpha$ and $\beta$, we can have a phase where the term $\left(6mK+\beta\right)\tau$ is very dominant over the term $\alpha\tau^2$ and thus obtain a decelerated power law expansion.

As a side remark, it should be noted that in the presence of potentials of the form \eqref{IFD1}, the Hamiltonian constraint \eqref{F2} reduces to a relation identical to \eqref{0.0.8}.  However now the integration constant $A$ does not have to satisfy the condition $A\geq 0$. Thus in this case no restrictions are imposed on the coupling constant $\epsilon$.

\subsubsection{A note on renormalizability in the case of a non trivial potential.}
So far, we have studied a list of potentials and we have given the expression of the single potential that condenses them all: altogether, they are capable of fitting within a unique scheme all expansion eras, but there is still a problem we must address about renormalizability. As it is well known, the presence of torsion renders the Dirac equation non-renormalizable; and as it is also widely recognized, non-minimal coupling do that too: one would then reasonably expect that torsion in non-minimal coupling would induce for the Dirac equation an even higher degree of non-renormalizability. But what happens is quite the contrary: opposite to our intuition, the degree of non-renormalizability is lowered. In fact, the resulting non-linear terms are even super-renormalizable \cite{FVC}. This is a nice result, and consequently it would be desirable that it be maintained also in presence of this potential. We split tho two cases: in the ultra-violet case, we have that
\begin{equation}
V(\bar\psi\psi\rightarrow\infty)\rightarrow-\frac{\beta}{6\epsilon K}
+\frac{\gamma\left(\bar\psi\psi\right)^{p}}{6\epsilon K^{p+1}\left(p-1\right)}
\end{equation}
so that the potential is reduced to one term that behave as a cosmological constant, which in high-energy physics is irrelevant, plus a term that scales as $\left(\bar\psi\psi\right)^{p}$, which therefore is renormalizable if and only if $p\leqslant\frac{4}{3}$, and specifically in the case of the equality the theory is renormalizable, while for the inequality the theory is super-renormalizable. In the infra-red case, it is 
\begin{equation}
V(\bar\psi\psi\rightarrow0)\rightarrow-\frac{\alpha}{6}
+\frac{\gamma\left(\bar\psi\psi\right)^{p+1}}{6K^{p+1}\left(p-1\right)}
\end{equation}
with a cosmological constant that now is relevant, and it constitutes the reason why the dark energy behaviour is recovered, plus an additional term in $\left(\bar\psi\psi\right)^{p+1}$, for which we have to require $p\geqslant-1$ if we want the results about dark energy preserved. All in all, the constraint given by $-1\leqslant p\leqslant\frac{4}{3}$ is the one that keeps the theory both in infra-red and in ultra-violet regimes completely renormalizable. And nicely, these are also the exact constraints we would need to get for $\frac{1-p}{2}=-\frac{1}{3q}$ the limiting condition $q\geq 2$ needed to provide inflation and also the limiting condition $p\geqslant-1$ needed to maintain the dark energy results. In this sense the potential we have furnished, together with the constraining conditions $-1\leqslant p\leqslant\frac{4}{3}$, is such that it recovers the correct dynamics for the expansion of the universe \emph{precisely} because it is the potential for which the theory is renormalizable. This is a 
surprisingly good double-take of the theory, because if at first the form of the potential might have looked quite arbitrary, and some might have thought it was chosen to yield the wanted cosmology, in reality that potential could not have been any different, or else the theory would have been ill-defined in terms of particle-physics. That the expected behaviour of the standard model of cosmology be implied by constraints on the standard model of particle-physics was, in our knowledge, not known before.

\subsection{In the presence of dust fluid}
In the case of presence of dust fluid with density $\rho$, the conservation laws for the fluid together with the relation $\tau=\frac{K}{\bar\psi\psi}$ ensure the relation
\begin{equation}\label{D1}
\rho = \frac{\rho_0}{\tau} = \frac{\rho_0}{K}\bar\psi\psi
\end{equation}
In such a circumstance, setting
\begin{equation}\label{D2}
\frac{\bar m}{2} := \frac{\rho_0}{K} + \frac{m}{2}
\end{equation}
it is easily seen that the dynamical equation for the scale volume $\tau$ becomes
\begin{equation}\label{D3}
2\frac{\ddot\tau}{\tau}\varphi+3\ddot\varphi+5\frac{\dot\tau}{\tau}\dot\varphi
=\frac{3\bar{m}K}{\tau\left(2-\frac{\epsilon K}{\tau}\right)} - \frac{3\left(\epsilon K + 2\tau\right)\/V}{\tau\left(2-\frac{\epsilon K}{\tau}\right)} + \frac{3\left(\epsilon\/K+\tau\right)KV'}{\tau^2\left(2-\frac{\epsilon\/K}{\tau}\right)}
\end{equation}
formally identical to eq. \eqref{1}, with $\bar m$ replacing $m$. The conclusion follows that, by substituting $m$ by $\bar m$, all results and conclusions stated in subsection A hold also in presence of dust. 


\subsection{In the presence of radiation fluid}
We consider the presence of a radiation fluid with equation of state $p=\frac{1}{3}\rho$. The conservation laws for the fluid provide the relation $\rho =\frac{\rho_0}{\tau^{\frac{4}{3}}}$.
In this case, the dynamical equation for $\tau$ is given by
\begin{equation}\label{R1}
2\frac{\ddot\tau}{\tau}\varphi+3\ddot\varphi+5\frac{\dot\tau}{\tau}\dot\varphi
=\frac{2\rho_0}{\tau^{\frac{4}{3}}} + \frac{3mK}{\tau\left(2-\frac{\epsilon K}{\tau}\right)} - \frac{3\left(\epsilon K + 2\tau\right)\/V}{\tau\left(2-\frac{\epsilon K}{\tau}\right)} + \frac{3\left(\epsilon\/K+\tau\right)KV'}{\tau^2\left(2-\frac{\epsilon\/K}{\tau}\right)}
\end{equation}
Choosing a potential of the form $V=\bar{V} + \tilde{V}$, with $\bar V$ satisfying the equation
\begin{equation}\label{R2}
\frac{2\rho_0}{\tau^{\frac{4}{3}}} - \frac{3\left(\epsilon K + 2\tau\right)\/\bar{V}}{\tau\left(2-\frac{\epsilon K}{\tau}\right)} + \frac{3\left(\epsilon\/K+\tau\right)K\bar{V}'}{\tau^2\left(2-\frac{\epsilon\/K}{\tau}\right)} = 0
\end{equation}
amounting to
\begin{equation}\label{R3}
\frac{2\rho_0}{K^{\frac{1}{3}}}(\bar\psi\psi)^{\frac{1}{3}} - \frac{3K\left(\epsilon\bar\psi\psi+2\right)}{\bar\psi\psi\left(2-\epsilon\bar\psi\psi\right)}\bar{V} + \frac{3K\left(\epsilon\bar\psi\psi+1\right)}{\left(2-\epsilon\bar\psi\psi\right)}\bar{V}' = 0,
\end{equation}
eq. \eqref{R1} reduces to 
\begin{equation}\label{R4}
2\frac{\ddot\tau}{\tau}\varphi+3\ddot\varphi+5\frac{\dot\tau}{\tau}\dot\varphi
=\frac{3mK}{\tau\left(2-\frac{\epsilon K}{\tau}\right)} - \frac{3\left(\epsilon K + 2\tau\right)\/\tilde{V}}{\tau\left(2-\frac{\epsilon K}{\tau}\right)} + \frac{3\left(\epsilon\/K+\tau\right)K\tilde{V}'}{\tau^2\left(2-\frac{\epsilon\/K}{\tau}\right)}
\end{equation}
which is identical to \eqref{1}. Again, an analysis analogous to that developed in subsection A is then applicable also in this case with identical results. A solution of \eqref{R3} is given by
\begin{equation}\label{R5}
\bar V = 2\rho_0\left(\frac{\bar\psi\psi}{K}\right)^{\frac{4}{3}}
\end{equation}


\subsubsection{The case $V=0$.}
In this case eq. \eqref{R1} simplifies to
\begin{equation}\label{R6}
2\frac{\ddot\tau}{\tau}\varphi+3\ddot\varphi+5\frac{\dot\tau}{\tau}\dot\varphi
=\frac{2\rho_0}{\tau^{\frac{4}{3}}} + \frac{3mK}{\tau\left(2-\frac{\epsilon K}{\tau}\right)} 
\end{equation}
which can be handled as above, giving rise to the final equation 
\begin{equation}\label{R7}
\left(2-\frac{\epsilon\/K}{\tau}\right)^2{\dot\tau}^2 = 12\rho_0\tau^{\frac{2}{3}} + \frac{12\epsilon\/K\rho_0}{\tau^{\frac{1}{3}}} + 6mK\tau + A
\end{equation}
$A$ being a suitable integration constant. For very small values of $\tau$ and supposing $\epsilon >0$, eq. \eqref{R7} can be approximated to
\begin{equation}\label{R8}
\epsilon\/K\frac{|\dot\tau|}{\tau} = \sqrt{12\epsilon\/K\rho_0}\tau^{-\frac{1}{6}}
\end{equation}
yielding $\tau \approx t^6$ which can account for an accelerated early phase of the universe (at least for isotropic models). {  As it is clear from \eqref{R8}, we underline that this dynamics is strictly due to the non--minimal coupling. This is a remarkable difference with respect to the minimally coupled theory where the presence of a fermionic self--interacting potential is necessary to generate inflationary phases at early time \cite{VFC}}. For very large values of $\tau$, eq. \eqref{R7} can be approximated to
\begin{equation}\label{R9}
2|\dot\tau| = \sqrt{6mK\tau}
\end{equation}
yielding $\tau \approx t^2$ and thus a decelerated expansion and isotropization of universe.


\subsubsection{Transition power law inflation -- decelerated power law expansion -- dark era}
Finally, taking the potential
\begin{equation}
V(\bar\psi\psi) = - \frac{\alpha}{6\left(\epsilon\bar\psi\psi+1\right)} - \frac{\beta\bar\psi\psi}{6K\left(\epsilon\bar\psi\psi+1\right)}
\end{equation}
into account, suitably choosing the parameters $\alpha$ and $\beta$ and repeating the arguments as in III.A.6, we recover again a phase transition: power law inflation -- decelerated power law expansion -- exponential expansion.


\section{Conclusions}
In this paper we have considered cosmological models in the framework of Einstein--Cartan--Sciama--Kibble gravity in which a Dirac field is non--minimally coupled to gravity. {  This non--minimal coupling has been investigated in a previous paper \cite{FVC} in connection with the renormalizability issue of Dirac equations. Here, we study some cosmological scenarios arising from such a theory}. In order to account for possible initial anisotropies of the universe, we have considered Bianchi--I models, looking at spatially flat FRW models as a particular case. We have shown that the non--minimal coupling can in general avoid the initial cosmological singularity in the scale volume (scale factors), in agreement with the results recently obtained in \cite{Magueijo:2012ug}, where another type of fermionic non-minimal coupling was studied. However this does not necessarily imply that the model is singularity free as the Hamiltonian constraint can induce bounds on $\epsilon$ which could cause a singularity in the 
shear at finite time. In this respect therefore, care should be taken in stating that these models are not ``singularity free''.

{  Using two different approaches, we have obtained several examples of fermionic self--interaction potential which generate a number of interesting cosmological phases (power law inflation, decelerated power law expansion, dark era). In fact by an accurate fine tuning,  even a transition  power law inflation --  decelerated power law expansion -- dark era is possible. Some of the potentials we obtained have the remarkable properties to be relatively simple combination of power of $\bar\psi\psi$ and to be able to lead dynamically to a dark era. The presence of cosmological fluids does not substantially modify the results achieved in the case only a Dirac field is present. We have analysed specifically the cases of dust and radiation. In this last case it became evident that the non minimal coupling alone is the origin of a power law inflation at early time.  

From our results it emerges that a fermionic self--interaction potential is necessary in order to generate an accelerated expansion phase of the universe at late time, when the contribution of the non--minimal coupling vanishes. Conversely and differently from what happens in the minimally coupled theory, in the presence of non--minimal coupling the fermionic potential can be no longer necessary for inflation; indeed there exist cases where the non--minimal coupling alone is sufficient to generate inflationary phases at early time (small values of scale volume). }

\begin{acknowledgments}
The authors would like to thanks the referee for useful comments that improved the original version of the manuscript.
\end{acknowledgments}

\end{document}